\documentstyle[epsfig,epsf,eqsecnum,aps,12pt]{revtex}
\def\beq{\begin{equation}}
\def\eeq{\end{equation}}

\def\bea{\arraycolsep .1em \begin{eqnarray}}
\def\eea{\end{eqnarray}}

\def\lsim{\mathrel{\lower4pt\hbox{$\sim$}}\hskip-12.5pt\raise1.6pt\hbox{$<$}\;}
\def\gsim{\mathrel{\lower4pt\hbox{$\sim$}}
\hskip-12.5pt\raise1.6pt\hbox{$>$}\;}

\def\s0#1#2{\mbox{\small{$ \frac{#1}{#2} $}}}
\def\0#1#2{\frac{#1}{#2}}

\begin{document}

\begin{center}

\thispagestyle{empty}

{\normalsize\begin{flushright}CERN-TH/2000-308\\[12ex] 
\end{flushright}}

\mbox{\large \bf 
Sum Rules in the CFL Phase of QCD at finite density}\\[6ex]

{Cristina Manuel
\footnote{E-Mail:  Cristina.Manuel@cern.ch}${}$ and Michel H.G. Tytgat
\footnote{E-Mail:  Michel.Tytgat@cern.ch}${}$}
\\[4ex]
{\it  Theory Division, CERN, CH-1211 Geneva 23, Switzerland.}
\\[10ex]
 
{\small \bf Abstract}\\[2ex]
\begin{minipage}[t]{14cm} 
We study the asymmetry between the vector current and axial-vector
current correlators in the colour-flavour locking 
(CFL) phase of QCD at finite density. Using Weinberg's sum rules,
we compute the decay 
constant $f_\pi$ of the Goldstone modes and find agreement with previous 
derivations. 
Using Das's sum rule, we also estimate the contribution of 
electromagnetic interactions to the mass of the charged modes.
Finally, we comment on low temperature corrections to the 
effective field theory describing the Goldstone bosons.
\end{minipage}
\end{center}

\newpage
\pagestyle{plain}
\setcounter{page}{1}

\section{Introduction}

The possibility that at large quark densities the ground state of QCD  is in a colour superconducting phase~\cite{Bailin:1984}
has attracted much attention recently (see {\em e.g.}
\cite{Schaefer:2000et,Rajagopal:2000uu}
for recent reviews). 
The pattern of symmetry breaking is most interesting for
 three light quark flavours. Alford, Rajagopal and Wilczek \cite{Alford:1998} have shown that the diquark condensates
 lock the colour and flavour 
symmetry transformations: in the chiral limit, this colour-flavour locking (CFL) condensate
 spontaneously breaks both colour and flavour symmetries
 $SU(3)_c \times SU(3)_L \times SU(3)_R \times U(1)_B
\rightarrow SU(3)_{c+L+R}$. 
Nine Goldstone modes survive to the Higgs 
 mechanism.~\footnote{If $U(1)_A$ breaking effects can be neglected,
there is  an extra Goldstone mode, analogous to the $\eta_0$.} One is a singlet, scalar mode, associated with 
 the breaking of $U(1)_B$.
 The other eight Goldstone bosons belong to an octet of 
 pseudoscalar modes associated with $SU(3)_L \times SU(3)_R$ chiral 
 symmetry breaking, just like in vacuum. At low energies, $E \ll \Delta$ where $\Delta$ is the gap, an effective lagrangian and
chiral perturbation theory ($\chi$PT) \cite{Gasser:1985gg}
can be used to study the dynamics of these light modes. Moreover, because of asymptotic freedom, at very large densities
  one can match the predictions of the effective theory
 with those of the underlying microscopic theory. 
 
There has been already quite some work on the effective theory approach
\cite{Casalbuoni:1999wu,Son:2000cm,Rho:2000xf,Hong:2000ei,Manuel:2000wm,Zarembo:2000pj,Beane:2000ms}.
In the present letter we will be concern with the  application of sum rules in the CFL phase. In particular, we will derive 
the expression for the pion decay constant as a direct consequence of chiral symmetry breaking. 
Although our result is not new, the formulation in term of sum rules is more
general and also provides a self-consistent check of previous results.
 We also estimate the effects of chiral symmetry breaking by 
electromagnetic interactions. 
    
\section{Weinberg  sum rules in the CFL phase of QCD}

We follow Casualboni and Gatto to construct the effective lagrangian~\cite{Casalbuoni:1999wu}.
With the diquark condensates defined as $X \sim \langle \psi_L \psi_L \rangle$ and 
$Y \sim \langle \psi_R \psi_R \rangle$, the low energy lagrangian in the chiral limit  
 is~\footnote{We have omitted  the Goldstone mode associated to the spontaneous
breaking of  $U(1)_B$.}
\bea
\label{lagran1}
{\cal L} & = & {\cal L}_V +  {\cal L}_A \\
         & = & \frac{f^2_\pi}{4}\, 
\left( {\rm Tr} \left( X \partial_0 X^\dagger
- Y \partial_0 Y^\dagger \right)^2 - v_\pi^2  {\rm Tr}
 \left( X \partial_i X^\dagger- Y \partial_i Y^\dagger \right)^2 \right) 
\nonumber \\
& + & \frac{a f^2_\pi}{4}\, \left( {\rm Tr} \left( X \partial_0 X^\dagger
+ Y \partial_0 Y^\dagger + 2 i g G_{0} ^a \frac{\lambda^a}{2} \right)^2 - b v^2_\pi
 {\rm Tr} \left( X \partial_i X^\dagger
+ Y \partial_i Y^\dagger + 2 i g G_{i} ^a \frac{\lambda^a}{2}\right)^2 \right)
\nonumber
\eea
where $G^a_\mu$
is the gluon field and $g$ is the gauge coupling constant.
Defining the colour singlet matrix $\Sigma = X Y^\dagger$,
the first term in Eq. (\ref{lagran1}) becomes 
\beq
\label{lagran2}
{\cal L}_V   = \frac{f_\pi^2}{4} \left( 
{\rm Tr} \left( \partial_0 \Sigma
\partial_0 \Sigma^\dagger \right) - 
v^2_\pi {\rm Tr} \left( \partial_i \Sigma
\partial_i \Sigma^\dagger \right) \right) \ .
\eeq
The second term in Eq.~(\ref{lagran1}) contains a mass term for the gluons.
Since the gluons are heavy they can
be integrated out and, to leading order at low energy,
the effective lagrangian reduces to Eq.~(\ref{lagran2}).
Except for the breaking of Lorentz symmetry, 
the lagrangian~(\ref{lagran1}) corresponds to the hidden gauge symmetry 
version of QCD in vacuum\cite{Bando:1988br}. In that
case, the vector mesons play the role of the gauge fields
associated with the hidden gauge symmetry \cite{Bando:1988br}.

In Eq.~(\ref{lagran1}) there are four independent parameters,
$f_\pi$, $v_\pi$, $a$ and $b$. These parameters can be
computed from the microscopic theory.
 The Debye and Meissner masses from 
Eq.~(\ref{lagran1}) are given respectively by $m_D^2 = a g^2 f^2_\pi$ and 
$m^2_M =b v^2_\pi m^2_D$. These have been first computed by Son and Stephanov~\cite{Son:2000cm}
from the microscopic theory and matched to those of
 the effective theory to derive
\beq
\label{values}
f_\pi^2 = {21 - 8 \ln{ 2} \over 18} \, {\mu^2\over 2 \pi^2} \ , 
\qquad v_\pi = \frac{1}{\sqrt{3}}\,.
\eeq
They also argued that $a=b=1$ to leading order in the gauge coupling,
because those values could be only modified by higher loop effects.
Although their conclusion is correct, the argument has
a potential flaw. We will see that, because of infrared divergences in
 the CFL phase of QCD, the connection
between the loop and  gauge coupling expansions is lost.
For instance, it is not {\em a priori} guaranteed that a two-loop diagram is
suppressed with respect to a one-loop diagram at weak coupling.

\bigskip

In this letter we present an alternative derivation of $f_\pi$,
which turns into a consistency check of the results in Eq. (\ref{values}).
Our computation relies on defining $f_\pi$ as the constant which
parametrises the asymmetry  between two-point
vector and axial-vector correlators in the ground state of the CFL phase.
Since Lorentz symmetry is broken
there are two possible independent correlators which can be defined.
If $V^a_\mu = {\bar \psi} \gamma^\mu \frac{\lambda^a}{2} \psi$ and
$A^a_\mu = {\bar \psi} \gamma^\mu \gamma_5 \frac{\lambda^a}{2} \psi$
are the vector and axial-vector flavour currents, respectively, then
in the chiral limit $m_q =0$,
\begin{mathletters}
\label{SR1}
\bea
\label{SR1a}
f^2_\pi \delta^{ab} &  = & i \int d^4 x \, 
\left\{ \langle V_0^a(x) V_0^b(0) \rangle - \langle A_0^a(x) A_0^b(0) \rangle
\right\} \ , \\
\label{SR1b}
- v^2_{\pi} f^2_\pi \delta^{ab} &  = & i \int d^4 x \, 
\left\{ \langle V_i^a(x) V_i^b(0) \rangle - \langle A_i^a(x) A_i^b(0) \rangle
\right\} \ .
\eea
\end{mathletters}
$\!\!$These relations can be inferred from the fact that the 
axial-vector currents taken between the ground state and a pion state
give~\cite{Pisarski:1996mt} (see appendix B)
\beq
\label{matrix}
\langle 0 | A_a^0| \pi^b (p) \rangle = i f_\pi \delta^{ab} p^0 \ ,
\qquad \langle 0 | A_a^i| \pi^b (p) \rangle = i v^2_\pi f_\pi \delta^{ab} p^i \ .
\eeq

Prior to the discovery of QCD, Weinberg~\cite{Weinberg:1967kj}  derived a
set of relations between the moments of the spectral functions of
vector and axial-vector currents in vacuum (see {\em e.g.}~\cite{deRafael:1997ea} for a review). 
If one writes the correlators of vector and axial-vector currents in
terms of their spectral densities, we could generalise
the Weinberg sum rule~\cite{Weinberg:1967kj} to the CFL phase.
We will not write explicitly these relations here, but
just note that they are directly deduced from Eqs. (\ref{SR1}).

From Eqs. (\ref{SR1}) one sees that $f_\pi$ is the 
order parameter for spontaneous chiral symmetry breaking. If we compute 
the Debye mass independently and
take the ratio between $m^2_D/g^2 f^2_\pi$,
we get $a$. Similarly, the constant $b$ is computed as
the ratio $m^2_M/m_D^2 v^2_\pi$.
Note that in the hidden symmetry approach to vector
mesons, the value $a=2$ allows to implement vector meson dominance
in the chiral lagrangian \cite{Bando:1988br}.
 
\bigskip

We will first review the computation of the Debye and Meissner masses.
This computation has been done by several authors \cite{Son:2000cm,Rischke:2000ra} but, for completeness, we
repeat it here using the natural quark basis for the CFL phase. The
Nambu-Gorkov quark propagators are diagonal in the colour-flavour basis
while the Feynman rules for the vertices are non-diagonal (appendix A).
To compute the Debye and Meissner masses, one evaluates the
gluon polarization tensor as in Fig.1,
\bea
\label{polar}
\Pi^{\mu \nu}_{a b} (k) &  = &  - i \frac{g^2}{32} 
\sum_{B,C=1}^9 {\rm Tr}\left( \lambda^B \lambda^a \lambda^C \right) 
{\rm Tr}\left( \lambda^C \lambda^b \lambda^B \right) I_{BC}^{\mu \nu} (k) \\ 
& + &   i \frac{g^2}{32} 
\sum_{B,C=1}^9 {\rm Tr}\left( \lambda^B \lambda^a \lambda^C \right) 
{\rm Tr}\left( \lambda^B \lambda^b \lambda^C \right) R_{BC}^{\mu \nu} (k) \ ,
\nonumber
\eea
where 
\begin{mathletters}
\label{inte}
\bea
I_{BC}^{\mu \nu} (k) & = &  \sum_{e = \pm} \int \frac{d^4 q}{(2 \pi)^4}
{\rm Tr} \left( \gamma^\mu S^e_C(q) \gamma^\nu S_B^e(q-k) \right) \ , 
\\
R_{BC}^{\mu \nu} (k) &  = &  \sum_{e = \pm} \int \frac{d^4 q}{(2 \pi)^4}
{\rm Tr} \left( \gamma^\mu \Xi^e_C(q) \gamma^\nu \Xi_B^{-e}(q-k) \right) \ ,
\eea
\end{mathletters}
$\!\!$and $S_A^{\pm}/\Xi^{\pm}_A$ are the diagonal/off-diagonal
terms  in the Nambu-Gorkov matrix propagator (see Appendix A).

The calculation of Eq.~(\ref{polar}) gets simpler
by realizing that the quark propagators $S^{\pm}_A$ and $\Xi^{\pm}_A$
are the same for all the values of $A$, except $A=9$. The $U(3)$ traces of
Eq.~(\ref{polar}) can then be easily evaluated.\footnote{We neglect the effect of the sextet component of the condensate, 
which is
a valid approximation at very high densities.} To compute the Debye  
and Meissner masses, it is enough to evaluate the polarisation
tensor at $k=0$,  
\bea
\Pi_{ab}^{00}(k=0) & = &
 -i \frac{g^2}{12} \delta_{ab} \left[7 I_{11}^{t} (k=0)
+ 2 I^{t}_{19} (k=0) + 2 R^{t}_{11}(k=0) - 2
R^{t}_{19} (k=0)  \right] \ , \\
\Pi_{ab}^{ij}(k=0) & = &
 -i \frac{g^2}{12} \delta^{ij} \delta_{ab} \left[7 I_{11}^{s} (k=0)
+ 2 I^{s}_{19} (k=0) + 2 R^{s}_{11}(k=0) - 2
R^{s}_{19} (k=0)  \right] \ .
\eea
The superscript $t$ and $s$ above refers to, respectively, the purely temporal and purely spatial
components of the tensors defined in Eqs.~(\ref{inte}). In the static limit, $I^{0i}_{AB} = R^{0i}_{AB} = 0$, thus
$\Pi^{0i}_{ab} (k=0) =0$.
An explicit computation of the integrals in Eqs.~(\ref{inte}) gives, to leading
order,
\begin{mathletters}
\label{leadvalues}
\bea
I^{t}_{11} (k=0) & = & I^{t}_{19} (k=0) = 2 i \frac{\mu^2}{2 \pi^2} \ ,  \\
R^{t}_{11} (k=0) &  = &  - 2 i \frac{\mu^2}{2 \pi^2} \ , \qquad
R^{t}_{19} (k=0) =  i \frac{8}{3} \ln{2}\, \frac{\mu^2}{2 \pi^2} \ , \\ 
I^{s}_{11} (k=0) & = & I^{s}_{19} (k=0) = - \frac{2}{3} i \frac{\mu^2}{2 \pi^2} \ , \\
R^{s}_{11} (k=0) & = & - \frac{1}{3} R^{t}_{11} (k=0) \ ,
\qquad R^{s}_{19} (k=0) =  - \frac{1}{3} R^{t}_{19} (k=0) \ .
\eea
\end{mathletters}
$\!\!$It is maybe worth  noticing that in the static limit the integrals in
Eqs.~(\ref{inte}) for the temporal components are dominated
by quark-quark excitations. The antiquark-antiquark  contribution is 
suppressed and can be neglected. Also the contribution of 
quark-antiquark excitations is precisely vanishing. However, for
the spatial components, the quark-antiquark excitations do contribute to $I^{s}_{11/19} (k=0)$.
This fact has already been stressed in~\cite{Rischke:2000ra}\footnote{We have assumed that the so-called antigap 
gives a subdominant contribution in the above calculations.}.
The Debye and Meissner masses are finally given by
\beq
m_D^2 = g^2 \,{21 - 8 \ln{ 2} \over 18} \, {\mu^2\over 2 \pi^2} \ , \qquad
m_M^2 = \frac{1}{3}\,m_D^2\ ,
\eeq
expressions which agree with the results of \cite{Son:2000cm,Rischke:2000ra}.

\bigskip
We now turn to the evaluation of the correlators of vector
and axial-vector currents in QCD. 
The best strategy to compute them
is to define the generating functional 
$Z$ for the Green's functions of vector and axial-vector quark
currents in presence of external sources $v_\mu$ and $a_\mu$
\beq
e^{i Z[v,a]} = \langle T \exp{\left(i \int dx \, {\bar \psi} \gamma_\mu
\left( v^\mu + \gamma_5 a^\mu \right) \psi \right)} \rangle \ .
\eeq
Then functional derivatives of $Z$ with respect to the sources 
evaluated at $v=a=0$ generate
the desired correlators. 

Now we face a little puzzle. 
Since the leading order expression for $f_\pi^2$ as computed by Son and Stephanov 
does not depend on $g$, one would naively expect that the sum rule
Eq. (\ref{SR1}) is saturated by  one-loop diagrams, {\em i.e.} 
with no gluon lines, as in Fig.2. However, due to
the Dirac structure of the Nambu-Gorkov propagator,
it is easy to verify that at this order, the r.h.s. of Eq.~(\ref{SR1}) is precisely zero !
But the next-to-leading order contribution is {\em a priori} ${\cal O}(g^2)$, which seems to contradict the 
fact that $f_\pi^2$ is ${\cal O}(g^0)$.  
Actually, this naive power counting is spoiled by infrared divergences and we will see that the 
next-to-leading order contribution is indeed independent of the gauge coupling. 

The first non-vanishing contribution to the r.h.s. of Eq.~(\ref{SR1}) 
actually arises solely from the  two-loop diagram depicted in Fig.3. Although it is naively of
order $g^2$, it is easy to see that this dependence precisely cancels. The diagram is 
infrared divergent $\propto 1/k^2$ because of the gluon pole.
But in the CFL phase and in the limit of zero external momentum,
 the gluon pole is trivially regulated by 
either the Debye or the Meissner mass. In both case, 
$1/k^2 \rightarrow 1/g^2 \mu^2$. Taking into account the factor of $g^2$ from the gluon vertices,
 we conclude that the diagram is ${\cal O}(g^0)$, as desired. 
There are other diagrams with the same topology but they do not contribute.
First, if we replace an external vector current by an axial-vector 
one, the diagram vanishes because of the trace over $\gamma_5$. Furthermore, we will see below 
that the diagrams with normal quark propagators ({\em i.e.} diagonal components of the Nambu-Gorkov matrix of propagators) do also vanish. 

\bigskip

We  evaluate the diagram of Fig.3 by computing
each quark loop separately. In  each of them, 
there is both a gluon vertex and a flavour vector current vertex.
 The computation is superficially the same as for the one-loop  polarisation
tensor (\ref{polar}). The important difference is that the Feynman rules for these two vertices
are different because the gluon vertex changes
colour but leave the flavour indices untouched, while the flavour vector
current vertex does the opposite. This is clearly reflected in the
Feynman rules in the CFL basis given in  Appendix A.  
For each loop  we find
\beq
  \frac{g}{32} \
\sum_{B,C=1}^9 {\rm Tr}\left( \lambda^B \lambda^a \lambda^C \right) 
{\rm Tr}\left( \lambda^C \lambda^b \lambda^B \right) R_{BC}^{\mu \nu} (k) 
\eeq
which we evaluate in the limit $k \rightarrow 0$. We find
\begin{mathletters}
\label{nedeed}
\bea
& \delta_{ab} &\frac{g}{12} \left(  
 7 R^{t}_{11} (k=0) + 2 R^{t}_{19} (k=0)
\right) = - i \delta_{ab} \frac{m^2_D}{g} \ , \\
& \delta_{ab} & \delta^{ij}\frac{g}{12} 
\left(  7 R^{s}_{11} (k=0) + 2 R^{s}_{19} (k=0)
\right) = - i \delta_{ab}  \delta^{ij} \frac{m^2_M}{ g} \ ,
\eea
\end{mathletters}
$\!\!$where we have used the results obtained in 
Eqs.~(\ref{leadvalues}). A similar one-loop diagram with the 
diagonal terms of the Nambu-Gorkov propagator gives
\beq 
 - \frac{g}{32} 
\sum_{B,C=1}^9 {\rm Tr}\left( \lambda^B \lambda^a \lambda^C \right) 
{\rm Tr}\left( \lambda^B \lambda^b \lambda^C \right) I_{BC}^{\mu \nu} (k) \ ,
\eeq
which in the $k \rightarrow 0$ limit reduces to
\begin{mathletters}
\bea
& \delta_{ab} &\frac{g}{12} \left( 2 I^{t}_{11} (k=0) -2 I^{t}_{19}(k=0)  
\right) = 0 \ , \\
& \delta_{ab} & \delta^{ij}\frac{g}{12} 
\left( 2 I^{s}_{11} (k=0) -2 I^{s}_{19}(k=0) 
\right) = 0 \ ,
\eea
\end{mathletters}
$\!\!$so that it vanishes, as we mentioned earlier.
Only the pure superconducting loop, that is,
the one which arises from the off-diagonal terms of the Nambu-Gorkov
propagator, contributes to the sum rule (\ref{SR1})
to leading order.

With the  results in Eqs. (\ref{nedeed}),
we easily get the expression of the
two-loop diagram of Fig.3.  Since the correlator in the sum rule
is evaluated at $k=0$, the gluon propagator is dominated by the
Debye (or Meissner) mass. It is then trivial to verify that
\bea
\Pi_{VV,ab}^{00} (0) -\Pi_{AA,ab}^{00} (0)
 & = &  i \delta_{ab} \frac{ m^2_D}{g} \, \frac{i \delta_{cd}}{m^2_D}  
\, i \delta_{db}\frac{ m^2_D}{g} =
- i  \delta_{ab} \frac{m^2_D}{g^2} \ , \\
\Pi_{VV,ab}^{ii} (0) -\Pi_{AA,ab}^{ii} (0)
 & = &  i \delta_{ab} \delta^{ik} \frac{ m^2_M}{g} \, \frac{-i 
\delta_{cd} \delta^{kl}}{m^2_M}  
\, i \delta_{db}\delta^{li} \frac{ m^2_M}{g} =
 i  \delta_{ab} \delta^{ii} \frac{m^2_M}{g^2} \ .
\eea
The Weinberg sum rules~(\ref{SR1}) finally give $f_\pi^2$ and $v_\pi$ as in Eq.~(\ref{values}). Incidentally, $a=b=1$ to leading order in the gauge 
coupling. 

We have thus checked that Eq.~(\ref{SR1}) is saturated to ${\cal O}(g^0)$ by the diagram of Fig.3. 
Clearly, {\em any} other diagram with one or more gluon lines is suppressed at 
weak coupling. 

\section{Sum rule with electromagnetic interactions}

The inclusion of electromagnetic interactions in the effective
low energy lagrangian is straightforward. One only has to
use the adequate covariant derivatives. Like in the Standard Model, 
one linear combination of the gluon and  photon fields become massive,
while the orthogonal one stays massless~\cite{Alford:1998}.
Integrating out the heavy gauge fields leaves us with
\beq
\label{lagrele}
{\cal L}    =   -\frac14 {\tilde F}^{\mu \nu} {\tilde F}_{\mu \nu} +
\frac{f_\pi^2}{4} \left( 
{\rm Tr} \left( D_0 \Sigma
D_0 \Sigma^\dagger \right) - 
v^2_\pi {\rm Tr} \left( D_i \Sigma
D_i \Sigma^\dagger \right) \right)  
 +  C {\rm Tr} \left( Q \Sigma Q \Sigma^\dagger \right)
 \ ,
\eeq
where 
\beq
D_\mu \Sigma = \partial_\mu \Sigma - i Q {\tilde A}_\mu \Sigma 
+ i \Sigma Q {\tilde A}_\mu \ , \qquad
 Q = {\tilde e}\, {\rm diag}(\frac23, -\frac13, -\frac13) \ ,
\eeq
and ${\tilde A}_\mu$ is the modified photon, while 
${\tilde e}$ is the effective coupling of the charged
Goldstone bosons with  the modified photon.
The last term in Eq.~(\ref{lagrele}) is  
allowed by the symmetries of the problem
and represents an explicit breaking of chiral symmetry
by the electromagnetic interactions ~\cite{Ecker:1989te}. 
This is seen as follows. The charge matrix $Q$ represents
an explicit chiral symmetry
breaking term in the QCD lagrangian. In order
to see its effects in the low energy lagrangian,
 one treats $Q$ as a spurion field whose vacuum expectation value
explicitly breaks the chiral symmetry.
To  restore the symmetry  one  has to introduce  left and
 right charge matrices which transform as $Q_L \rightarrow U_L Q_L U_L^\dagger$,
and $Q_R \rightarrow U_R Q_R U_R^\dagger$ if the meson field matrix
transforms as $\Sigma \rightarrow U_L \Sigma U_R ^\dagger$. The 
 term ${\rm Tr} \left( Q_L \Sigma Q_R \Sigma^\dagger \right)$
is thus allowed in the effective lagrangian. The terms
 ${\rm Tr}(Q_L^2)$ and ${\rm Tr}(Q_R^2)$ are also allowed, but these
only represent shifts in the ground state energy due to electromagnetism.

To leading order in the meson fields,
the last term of Eq. (\ref{lagrele})
represents the contribution of electromagnetic interactions to
the masses of the charged Goldstone bosons.
In particular, using the standard parametrisation of $\Sigma$, one finds in the
chiral limit 
\beq
M^2_{\pi^\pm} = M^2_{K^\pm} = \frac{2 {\tilde e}^2 C}{f^2_\pi}  \ .
\eeq
In the chiral limit the constant $C$ obeys a sum rule,
which was first derived in the context of QCD in vacuum in \cite{Ecker:1989te},
\beq
\label{SR2}
C \delta^{ab} = \frac12 \int d^4x\, D^{\mu \nu}(x) 
\left\{ \langle V_\mu^a(x) V_\nu^b(0) \rangle - 
\langle A_\mu^a(x) A_\nu^b(0) \rangle
\right\} \ ,
\eeq
where $D_{\mu \nu}$ is the photon propagator. In momentum space
\beq
\label{momenSR2}
C \delta^{ab} = \frac12 \int \frac{d^4 k}{(2 \pi)^4}\, D_{\mu \nu}(k) 
\left\{ \Pi_{VV}^{\mu \nu, ab} (k) - \Pi_{AA}^{\mu \nu, ab} (k) 
\right\} \ .
\eeq
This sum rule was derived prior to the discovery
of QCD using quite general
assumptions based on soft pion theorems and current
algebra~\cite{Das:1967it}. 
The sum rule~(\ref{SR2}) gives a
gauge independent value of $C$ thanks to vector and axial-vector currents conservation.  
Furthermore, using Weinberg sum
rules, one can check that the expression is ultraviolet finite.
This is because
these sum rules are essentially a statement about the
convergence of the difference between vector and axial vector correlators
for large Euclidean momentum (see {\em e.g.}~\cite{deRafael:1997ea}).
It is also possible to see how charged pion and kaon masses diminish 
with temperature because of the tendency towards chiral symmetry restoration~\cite{Manuel:1999sy}. 

In QCD in vacuum, using the Weinberg and Das sum rules
and the hypothesis of vector meson dominance, one gets a mass splitting between 
$\pi^\pm$ and $\pi^0$ which is in good agreement 
with experiments\footnote{For kaons the
sum rule does not work so well. In this case the
mass difference of kaons is dominated by  $m_s\not= 0$ effects.}.
One could apply the sum rule~(\ref{SR2}) to compute the contribution of electromagnetic interactions 
to the mass of charged Goldstone modes in the CFL phase.
An attempt in this direction has been made by
Hong~\cite{Hong:2000ng}. He considered a  diagram with  one quark-loop convoluted with a photon
propagator. Why this is 
incorrect  can be seen in two different ways. Firstly, from the sum rule~(\ref{SR2}). The same 
reasoning as in the previous section shows 
that the calculation must involve at least two quark loops: at the one quark loop level,
 the r.h.s of Eq.(\ref{SR2})
simply vanishes. 
Secondly, from the effective lagrangian~(\ref{lagrele}). Because 
$\Sigma \rightarrow U_L \Sigma U_R^\dagger$ under chiral transformations, the operator ${\rm Tr} 
\left( Q \Sigma Q \Sigma^\dagger \right)$ involves both quark chiralities. In the CFL phase, this is only 
possible with two quark loops.  At the one quark loop level, electromagnetic interactions merely shift the ground state energy, an effect which in the effective lagrangian can be represented by a term
  $\propto {\rm Tr} (Q^2)$.

The leading contribution to the sum rule (\ref{SR2}) comes
from the diagram of Fig.4. However, compared to the previous section, the 
difference between the vector and axial-vector correlators has to be computed
at non-zero momentum. This substantially complicates calculations
as the quark, gluon and photon propagators have to be known over
quite distinct regimes, $0 \lsim k \lsim \Delta$, $\Delta \lsim k \lsim \mu$ and $\mu \lsim k$. 
We can however estimate Eq.(\ref{SR2}) if we
assume that the integral in Eq. (\ref{momenSR2}) is restricted to the 
range $0 \lsim k \lsim \Delta$ because $R^{\mu \nu}(k) \rightarrow 0$ for $k \gg \Delta$.
 If we furthermore take the gap 
to be constant over this range, 
we get 
\beq
\label{bulshit}
M^2_{\pi^\pm} = M^2_{K^\pm} \sim  {\tilde e}^2  \Delta^2 \ .
\eeq
This naive estimate is in rough agreement with the result
of Hong~\cite{Hong:2000ng} 
$M^2_{\pi^\pm} \sim {\tilde e}^2  \Delta^2 \log {\mu\over \Delta}$.
We were  
not able to deduce  whether our estimate could be spoiled  by the appearance of log-factors, 
as in Hong's expression. Obviously, it would  be interesting 
to perform the explicit calculation. 
In this regard, it is maybe worth noting that, 
contrary to what happens for finite 
quark mass effects, the electromagnetic mass of the charged mesons does not
tend to zero at very large densities.

Finally, we would like to comment on the behaviour of the mass
of the charged mesons at low temperature $T$. These can
be easily obtained from the analysis performed in \cite{Manuel:1999sy}
for the chiral lagrangian in vacuum, simply taking into account 
that the velocity of the Goldstone modes is now $v_\pi$.

Low temperature corrections to the value of the diquark condensate
have been studied in \cite{Casalbuoni:1999zi}. For temperatures
$T \ll f_\pi$, and at leading order, one finds that the pion decay
constant diminishes, while the velocity of the Goldstone modes
remains unchanged
\beq
f_\pi (T) = f_\pi \left(1 - \frac{1}{8} \frac{T^2}{v^3_\pi f^2_\pi} \right)
\ , \qquad v_\pi (T) = v_\pi \ .
\eeq
Like for QCD at zero chemical potential and low temperatures
\cite{Pisarski:1996mt}, we expect that the velocity of the Goldstone
modes is modified at order $T^4/f^4_\pi$.

The mass of the charged mesons at low temperature is \cite{Manuel:1999sy}
\beq
M^2_{\pi^\pm} (T) = M^2_{K^\pm} (T) = \frac{ 2 {\tilde e}^2 C(T)} {f^2_\pi(T)} \ ,
\eeq
where
\beq
C(T) = C \left( 1 - \frac{1}{2} \frac{T^2}{v^3_\pi f^2_\pi} \right)
+ \frac{ {\tilde e}^2 T^2}{4 v^3_\pi} \ ,
\eeq
so that 
\beq
M^2_{\pi^\pm} (T)  \approx  \left( 1 - \frac{1}{4} \frac{T^2}{v^3_\pi f^2_\pi} \right)M^2_{\pi^\pm} 
+ \frac{{\tilde e}^2 T^2}{4 v^3_\pi} \ .
\eeq
Without the explicit value of the zero temperature mass, it is however
not possible to determine whether the value of this mass diminishes
fast or not. In any case, these corrections are only valid for low temperature $ T \ll T_c \sim \Delta$. 
At the critical temperature $T_c$, the Cooper condensates melt and all the
broken symmetries in the system are restored.

\bigskip 

\bigskip

{\bf Acknowledgements:}
C.M. would like to thank Antonio Pich for useful discussions.
M.T. thanks Thomas Sch\" afer and Misha Stephanov
for stimulating conversations. C. M. acknowledges financial support
from a Marie Curie EC Grant (HPMF-CT-1999-00391).

\appendix

\section{Feynman Rules in the CFL basis}

Due to the complex colour-flavour structure of the condensates in the
CFL phase, it proves convenient to work in a quark basis in which the
quark propagator is diagonal, while the Feynman rules for the vertices are
non-diagonal~\cite{Son:2000cm}.

We define the colour-flavour basis by the transformation

\beq
\psi_{ai} = \frac{1}{\sqrt{2}} \sum_{A=1}^9 \lambda_{ai}^A \psi^A \ ,
\eeq
where $a$ and $i$ refer to the quark flavour and colour indices, respectively,
and $\lambda^A$, for $A=1, \dots,8$ are the Gell-Mann matrices, while
$\lambda^9 = \sqrt{\frac 23} 1$. 
In this basis the gap matrix diagonalizes
\beq
\Delta^{AB} = \delta^{AB} \Delta_A \, {\rm diag} (1,-1,1,1,-1,1,-1,1,-2) \ .
\eeq
We have neglected above the effect of the sextet.
The Nambu-Gorkov matrix propagator also becomes diagonal
in the $A$ index
\beq
S_{AB} = i \delta_{AB} \,\left(
\begin{array}{cc}
S_A^+ & \Xi^-_A \\
\Xi^+_A & S_A^- 
\end{array} \right) \ ,
\eeq
where
\bea
S_A^{\pm} (q) & = &\frac{\Lambda_{\bf q}^{\pm} \gamma^0 \left(q_0 \mp \mu \pm |{\bf q}| \right)}
{q_0^2 - \epsilon_A^2}  +
\frac{\Lambda_{\bf q}^{\mp} \gamma^0 \left(q_0 \mp \mu \mp |{\bf q}| \right)}
{q_0^2 - {\bar \epsilon}_A^2}  \ , \\
\Xi_A^- (q) & = & \gamma_5 \left\{ \frac{\Lambda_{\bf q}^- \Delta_A}
{q_0^2 - \epsilon_A^2}  + \frac{\Lambda_{\bf q}^+ {\bar \Delta}_A}
{q_0^2 - {\bar \epsilon}_A^2}  
\right\} \ , \\
\Xi_A^+ (q) & = & - \gamma_5 \left\{ \frac{\Lambda_{\bf q}^+ \Delta^*_A}
{q_0^2 - \epsilon_A^2}  + \frac{\Lambda_{\bf q}^- {\bar \Delta}^*_A}
{q_0^2 - {\bar \epsilon}_A^2}  
\right\} \ , 
\eea
where 
\beq
\Lambda_{\bf q} ^{\pm} = \frac{1    \pm  \gamma_ 0 {\bf \gamma} \cdot
{\bf \hat q}  }{2 } \ .
\eeq
are the positive/negative energy projectors, and
\beq
\epsilon_A = \sqrt{ (\mu - |{\bf q}|)^2 + \Delta^2_A} \ , \qquad
{\bar \epsilon}_A = \sqrt{ (\mu + |{\bf q}|)^2 + {\bar \Delta}^2_A} \ ,
\eeq
where $\Delta_A$ and ${\bar \Delta}_A$ are the gap and the antigap,
respectively. 

In the CFL basis, the fermion-gluon interaction term becomes
\beq
\label{gluonver}
\frac{g}{4} \sum_{B,C=1}^9 \sum_{a=1}^8 
{\rm Tr}\left( \lambda^B \lambda^a \lambda^C \right)
{\bar \psi}^B \gamma^\mu G_\mu^a \psi^C \ .
\eeq

In the CFL basis, the coupling of the fermions with external vector
and axial-vector sources reads
\beq
\label{flavorver}
\frac{1}{4} \sum_{B,C=1}^9 \sum_{a=1}^8 
{\rm Tr}\left( \lambda^C \lambda^a \lambda^B \right)
{\bar \psi}^B \gamma^\mu \left( v_\mu^a + \gamma_5 a_\mu^a \right) \psi^C \ .
\eeq

With the terms (\ref{gluonver}) and (\ref{flavorver}) one can immediately
deduce the Feynman rules for these vertices for the quarks fields and
the charge conjugate ones.

\section{Lorentz structure of Weinberg sum rules}

An elegant way to derive the sum rule ~(\ref{SR1}) is to start from the effective lagrangian~(\ref{lagran1}) and to introduce 
external vector and axial-vector local fields~\cite{Gasser:1985gg},
\beq
\partial_\mu \Sigma \rightarrow D_\mu \Sigma = \partial_\mu \Sigma - i (v_\mu + a_\mu) \Sigma + i \Sigma (v_\mu - a_\mu) \ .
\eeq
In a phase with spontaneously broken chiral symmetry, only the
axial-axial current correlator receives a  contribution from single 
Goldstone mode exchange.
To get its contribution to 
the axial-axial correlator, we thus set $v_\mu$ to zero, expand $\Sigma = \exp(i \phi/f_\pi)$ 
to second order in the meson field $\phi = \phi^a \lambda^a$, and keep the quadratic part of the effective lagrangian,
\beq
\label{quadratic}
{\cal L}_2 \approx {1\over 2} ( \partial_0 \phi^a - f_\pi a^a_0)^2 - {1\over 2} v_\pi^2\,(\partial_i \phi^a - f_\pi a_i^a)^2 \ .
\eeq
Finally, integrating over the $\phi^a$ field in the path integral to get the (mean field) expression for the
axial-vector current correlator, 
\beq
\exp i Z[a_\mu] = \int {\cal D}\phi \exp({i\! \int \!{\cal L}_2 }) \approx
 \exp \left\{ \frac{i}{2} \int\! d^4x \, d^4y
 \,a_\mu^a(x) \langle A_\mu^a(x) A_\nu^b(y) \rangle a_\nu^b(y) \right\}
\eeq
gives, in momentum space,
\beq
\langle A^a_0 A^b_0 \rangle  = i f_\pi^2 \, \delta^{ab} \left \{ g_{00} - {p_0^2 \over p_0^2 - v_\pi^2\, {\bf p}^2} \right \},
\eeq
\beq
\langle A^a_i A^b_j \rangle  = i f_\pi^2 v_\pi^2\, \delta^{ab} \left \{ g_{ij} - {p_i p_j v_\pi^2\over p_0^2 - v_\pi^2 \, {\bf p}^2} \right \} \ ,
\eeq
and
\beq
\langle A^a_0 A^b_i \rangle = - i f_\pi^2 v_\pi^2 \, \delta^{ab}
 { p_0 p_i \over p_0^2 - v_\pi^2 \, {\bf p}^2} \ .
\eeq
One can verify that the correlator is transverse, $P^\mu \langle A_\mu A_\nu\rangle =0$. For instance, in the usual basis of transverse operators,
\beq
P_{\mu\nu}^L = - (g_{\mu\nu} - P_\mu P_\nu/P^2) - P_{\mu\nu}^T,
\eeq
with
\beq
P_{ij}^T = \delta_{ij} - {p_i p_j\over {\bf p}^2}
\eeq
and $P_{\mu\nu}^T=0$ otherwise, the correlator reads 
\beq
\langle A_\mu^a A_\nu^b \rangle = - i f_\pi^2 v_\pi^2 \,\delta^{ab}
\left ( {P^2\over p_0^2 - v_\pi^2 \,{\bf p}^2} P_{\mu\nu}^L + P_{\mu\nu}^T\right)
\eeq
Except for the contact terms, the correlator can also 
be deduced by saturating it with a single intermediate 
meson and using the
matrix elements (\ref{matrix}). The contact terms can then be fixed by
 imposing current conservation.


\newpage

\begin{figure}[h]
\begin{center}
\epsfig{file=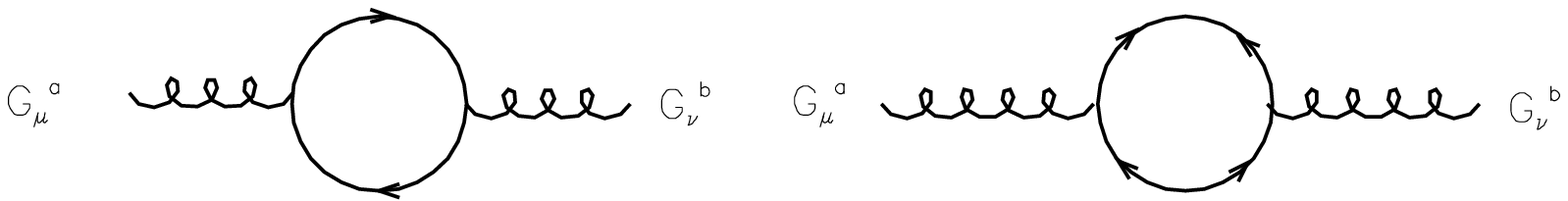,height=2cm}
\vskip 0.5cm
\caption{Gluon polarization tensor in the  CFL phase.}
\end{center}
\end{figure}

\bigskip

\begin{figure}[h]
\begin{center}
\epsfig{file=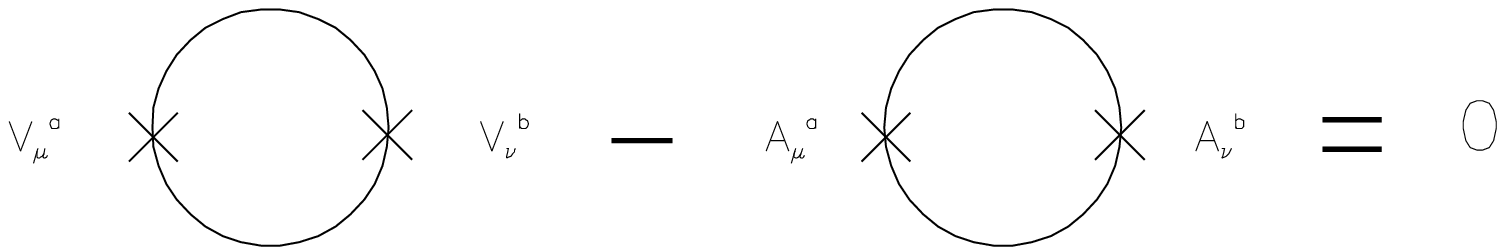,height=2cm}
\vskip 0.5cm
\caption{Sum rule at one-quark loop order. The quark lines are Nambu-Gorkov propagators. The crosses represent insertions of external flavour vector or axial-vector currents.}
\end{center}
\end{figure}
\bigskip

\begin{figure}[h]
\begin{center}
\label{sumrule1}
\epsfig{file=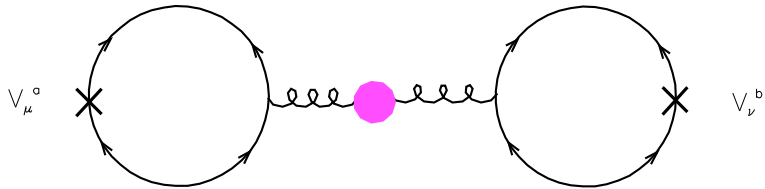,height=2.5cm}
\vskip 0.5cm
\caption{Leading order contribution to Eqs. (\ref{SR1}). The  dot represents
 the  resummation effects in the
gluon propagator.}
\end{center}
\end{figure}
\bigskip

\begin{figure}[h]
\begin{center}
\label{sumrule2}
\epsfig{file=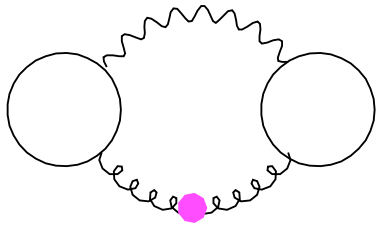,height=3cm}
\vskip 0.5cm
\caption{Leading order contribution to the electromagnetic mass of charged 
mesons. The wavy line corresponds to the photon propagator.}
\end{center}
\end{figure}

\end{document}